# Using Stereoscopic 3D Technologies for the Diagnosis and Treatment of Amblyopia in Children

Angelo Gargantini

Dipartimento di Ingegneria dell'Informazione e Metodi Matematici

Università di Bergamo - Italy

e-mail: angelo.gargantini@unibg.it

http://cs.unibg.it/gargantini

*Abstract*—The **3D4Amb** project aims at developing a system based on the stereoscopic 3D techonlogy. like the NVIDIA 3D Vision, for the diagnosis and treatment of amblyopia in young children. It exploits the active shutter technology to provide binocular vision, i.e. to show different images to the amblyotic (or lazy) and the normal eye. It would allow easy diagnosis of amblyopia and its treatment by means of interactive games or other entertainment activities. It should not suffer from the compliance problems of the classical treatment, it is suitable to domestic use, and it could at least partially substitute occlusion or patching of the normal eye.

## I. Introduction

*MBLYOPIA*, otherwise known as 'lazy eye', is reduced $A$ visual acuity that results in poor or indistinct vision in an eye that is otherwise physically normal, or out of proportion to associated structural abnormalities. It may exist even in the absence of any detectable organic disease. Typically amblyopia is present in only one eye and is generally associated with a squint or unequal lenses in the prescription spectacles. This low vision is not correctable (or only partially) by glasses or contact lenses.

There exist several causes of amblyopia. Anything that interferes with clear vision in either eye during the critical period (birth to 6 years of age) can result in amblyopia. The most common causes of amblyopia are constant strabismus (constant turn of one eye), anisometropia (different vision/prescriptions in each eye), and/or blockage of an eye due to trauma, lid droop, etc. If one eye sees clearly and the other sees a blur, the good eye and brain will inhibit

This work is partially supported by NVIDIA corp. under the Professor Partnerships program.
For further information see http://3d4amb.unibg.it

the eye with the blur. The brain, for some reason, does not fully acknowledge the images seen by the amblyopic or lazy eye. Thus, amblyopia is a neurologically active process. The inhibition process (*suppression*) can result in a permanent decrease of the vision in that eye that can not be corrected with glasses, lenses, or surgery. This condition affects 2-3% of the population, which equates to conservatively around 10 million people under the age of 8 years worldwide. Children who are not successfully treated when still young (generally before the age of 7) will become amblyotic adults. As amblyotic adults, they will have a normal life, except that they are prohibited from some occupations and they are exposed to a higher risk of losing the good eye due to injury or eye disease and became seriously visually impaired.

Amblyopia is currently treated by wearing an adhesive patch over the non-amblyopic eye for several hours per day, over a period of several months. This treatment can require up to 400 hours in total to be effective [3]. This conventional patching or occlusion treatment for amblyopia often gives disappointing results for several reasons: it is unpopular, prolonged, and it can sometimes make the squint worse because it disrupts whatever fusion there is. These issues frequently results in poor or *non-compliance* and since the success of patching depends on compliance, it performs on average very poorly. The treatment by itself works well, but it is often abandoned because it is too much trouble to take. Very often, children are averse to wearing a patch and parents found occlusion difficult to implement [5]. As noted in [12], a treatment whose unacceptability is greater than the motivation of the patients to apply it, will be often abandoned. And if the treatment of patching is not continued, it will eventually fail [11]. For this reason, the orthoptists and ophthalmologists are continuously looking for a more acceptable solution to the problem, i.e. an effective treatment that is also complied with and so really works [8].

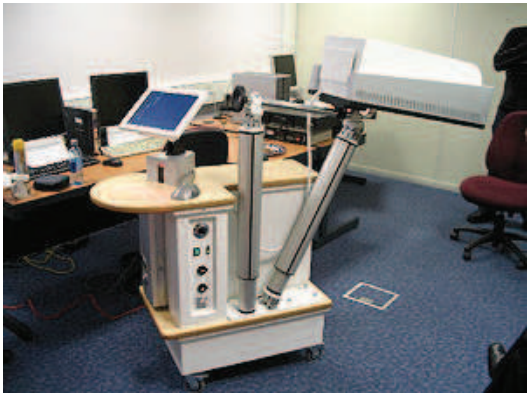

Figure 1: The I-BiT™ system

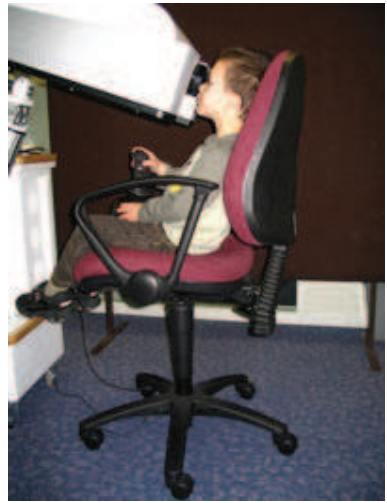

Figure 2: Playing with I-BiT™

## A. Computer based treatment of amblyopia

In the last years, several research groups have experimented treatment of amblyopia by exploiting and adapting information technologies. There exist several Personal Computer (PC) based software programs that allow vision training but they still require patching of the good eye. The most original system which does not require patching, is described in the following.

The VIRART group at the University of Nottingham has developed a novel virtual-reality (VR)-based display system which avoids occlusion of the nonamblyopic eye and facilitates the treatment of amblyopia with both eyes stimulated simultaneously [6]. This system is called I-BiT™. The I-BiT™ system itself consists of a viewer which is linked to a PC. The PC has a standard monitor for the clinician while the viewer allows binocular vision. Several types of viewers were tested, like binocular headsets and several types of viewers called "cyberscopes" similar to that shown in Figure 1.

This system incorporates adapted VR technology and specially written software providing interactive 2D and 3D games and videos to the patient via a stereo (binocular) display.

Treatment consists of watching video clips and playing interactive games in the clinic with specifically designed software to allow streamed binocular image presentation. The children sit in front of the viewer and play with the software designed for this kind of treatment, as shown in Figure 2.

*a) An evaluation of I-BiT™:* The experiments show that this type of treatment can be efficiently employed and it performs better than the classical treatment. In [15], the results of the use of I-BiT™ in six children are presented. In the case study, treatment consisted of watching video clips and playing simple interactive games with specifically designed software to allow streamed binocular image presentation via I-BiT™. Improvements in vision were demonstrable within a short period of time, in some children after 1h of treatment. In another case study [4], the system was applied in 12 older amblyopes (from 6 to 11 years), who had not complied with or responded to occlusion. Virtual reality images were projected to each eye simultaneously via a headset during eight treatment sessions of 25-min duration. Improvements in the vision were observed in more than half of the children.

However, the proposed treatment has some limits. The kind of hardware system used makes the treatment rather costly and performable only in suitable clinic rooms under the supervision of a doctor (or at least of an adult). The cure can be performed only for a limited time and only with precise time scheduling. For these reasons, we believe that the I-BiT™ system suffers from the same problem of compliance of the patching treatment. The goal of this project is to design a system which could combine the performance of I-BiT™ but be more accessible and usable.

*1) Other Binocular Computer-based Approaches:* Other research works present similar approaches. In [7], the authors describe their preliminary efforts to develop a convenient and viable binocular head mounted display (HMD) interface to rebalance the vision between the good and the lazy eye. The image presented to the normal eye will be attenuated while the image presented to the amblyopic eye will be enhanced. In [13] the authors present a multimedia viewer that uses high contrast images of increasing complexity to stimulate the lazy eye. In [9], the authors developed a software system to examine and to treat squint (small and middle range) and amblyopia. It requires an LCD display and anaglyph glasses.

At the best of our knowledge, no group has experimented the use of the stereo 3D vision systems for the diagnose and treatment of the amblyopia.

In this paper we present a project, 3D4AMB which is based on the 3D vision, for the diagnosis and treatment of amblyopia in young children. Section II introduces the basic principles of the system we have devised, while Section III presents several use cases we have modelled for this system. Section IV reports the software architecture.

## II. USING THE 3D TECHNOLOGY FOR BINOCULAR VISION

The main goal of this research project, called 3D4AMB, is to develop a system for the diagnosis and treatment of amblyopia, based on the binocular vision but that is accessible. With the term "accessible" we mean:

- **Inexpensive**: The system needs to be relatively *low* in cost, it must be affordable by a family. To be so cheap, the system may be based on standard off the shelf technologies, which could be bought in stores open to the general public.
- **Friendly to use**: The system needs to be friendly in its use such that the patients can use it without requiring a particular education or skill. The system may be operated autonomously by the children themselves and the intervention of an adult may be limited to initially set up the system (installation) and to start the treatment at least. To be so easily usable, the system may use standard user interfaces like a joystick, a remote control, or a mouse.
- **Suitable for domestic use**: The system can be used at home without frequent time-consuming visits to the hospital. In this way, the timing of the treatment can be decided by the patients (and the system must be able to record the actual use of it). It may use other domestic appliances normally used not for the treatment itself like standard personal computers and televisions.
- **Easily extensible**: It must be possible to easily develop new applications and programs to be added to the system. For this reason, standards and open software libraries may be used for developing the applications.

We have devised a system which has all the above characteristics and is based on the 3D technologies, although the goal is not to provide the patients with the 3D experience but to allow binocular vision. The classical use of a 3D system is to provide the two eyes with two different images of the same scene with a slightly offset viewing angles which correspond to the different viewpoints of our left and right eye. This vision produces an illusion of real depth of the scene and it is the basis of the *virtual reality*. We exploit only the capability of the 3D system to send two different images to the eyes while we do not want to recreate a virtual reality.

We have already built a working prototype based on the NVIDIA® 3D Vision™ technology, although other 3D technologies may be supported as well in the future. The NVIDIA 3D Vision technology is one of the most accessible 3D technologies available on the market today, it requires a standard personal computer with a NVIDIA graphic card (also entry level NVIDIA graphic boards work), a monitor 3D Vision ready, which is capable of a refresh rate of 120 Mhz, and a NVIDIA LCD shutter glasses. The NVIDIA 3D vision is based on active shutter technology. With this method, the left and right eye images are presented on alternating frames, but since these monitors are capable of 120Hz, each eye still sees a full 60Hz signal that is equivalent to the refresh rate on LCD monitors today. This offers a number of advantages with respect to other stereoscopic technologies like polarized or anaglyphs glasses or head-mounted displays, including:

- **Full image quality per eye**: In 3D mode, each eye receives the full resolution of the display for the highest possible image quality for text and objects. The colors are not altered and both eyes can receive images of the same quality.
- **Wide viewing angle for 3D**: Because full images are presented on alternating frames, there are no restrictions on the viewing angle in 3D mode. Users are free to move their heads vertically or horizontally within the full viewing angle of the display without losing the 3D effect or suffering increased ghosting which allowing for comfortable viewing for continuous gaming or movie watching.
- **Excellent 2D Operation**: If the user decides to switch back to 2D at any time (for the normal use of the PC), the performance of 3D Vision PCs support a 120 Hz higher refresh rate that reduces ghosting and blur typically found during motion on existing PCs that have 60 Hz displays.
- **Personalized fit**: The NVIDIA glasses have top-of-the-line optics, include adjustable nose pieces, and are modeled after modern sunglasses. They can be worn over prescription glasses.
- **Acceptable cost**: The cost of the glasses is around 140 USD, while the 3D ready monitors cost a little more than the traditional monitors but they offer a better quality also in 2D.

The system we have developed for 3D4Amb consists in a normal PC desktop connected to a 3D monitor (3D Vision-Ready Display). The PC must be 3D capable and have all the 3D4Amb software installed on it. The patient wears the NVIDIA active LCD shutter glasses that allow viewing a different image from the left and right eye. The scenario is depicted in Figure 3.

The basic principle of the system is that the amblyopic or 'lazy' and the normal eye are shown two different but related images. This principle can be used in practice for the treatment of amblyopia, where the amblyopic eye is shown the more interesting part of the images of the clip or of the game, while the non-amblyopic or 'good' eye is shown the less interesting part of the image. The content to be shown by the patient (game or image) is split by 3D4Amb in two parts, one for the right eye (the amblyopic eye in the Figure) and one for the left eye (the good eye in the Figure). The 3D4Amb software will decide what to send to both eyes depending on the type of treatment suggested by the physician. Note that the lazy eye of the child is more stimulated to work, but the non-amblyopic eye is not patched. The patient brain must join the two images to successfully see the complete image and successfully perform simple tasks in case of an interactive game. To make sure that the patient can join the two images there are a significant number of elements common to both images. Note that the final image is a bidimensional image because the goal in not to stimulate the stereo vision of the patient (at least initially).

## III. Use Cases of 3D4Amb

We have designed the following use cases that model the ways we expect 3D4Amb will be used and the possible interactions with physicians and patients. We have also developed simple prototype applications to prove that the designed usage of 3D4Amb is feasible.

*a) Diagnosis.:* The 3D4Amb system can be used for the screening and measurement of the amblyopia. The physician will save the parameters corresponding to the kind of amblyopia and these data will be reloaded by the 3D4Amb software used by the patient.

We have developed a simple program for the diagnosis. It takes an image or a geometrical shape (like a box) and splits it in two parts, one for each eye and check if the patient can merge again the two images into one image. Amblyotic kids may no be able to recognize the images. Another possible use of 3D4Amb is an application that shows images which amblyotic children can recognize but non amblyotic people cannot. This can be realized by taking a geometrical shape or an image and add some noise and split the images in two parts one of which contains more noise and the other much less noise. If the patient can still recognized the picture, that may be because the eye receiving more noise is lazy and the fusion is weak. Healthy patients with complete fusion are unable to recognize the images due the added noise.

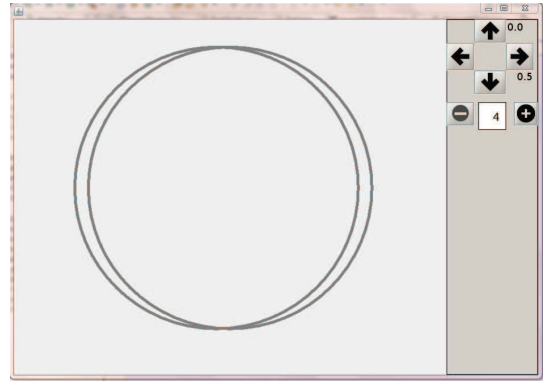

Figure 4: Diagnosis Application

It may be also used to measure convergence insufficiency in a similar way as stereograms and for an estimation of squint angle in the presence of strabismus. In Figure 4 we show a simple application that permits to measure the squint between the two eyes. Two circles are shown to the patient, one for each eye. The two circles are translated until the patient sees only one circle. This application is similar to that presented in [9].

Note that there are already computer based methods for screening children [10], but 3D4Amb should work much better: it should be more reliable and faster since it does require the occlusion of one eye.

*b) Passive Image and Movie Viewing.:* Another use of 3D4Amb is to visualize images and clips. 3D4Amb includes an image and video viewer that is able to send two different images to both eyes. This activity, although it is passive, since it does not require any action by the child, could be performed for prolonged time and would allow to exercise the lazy eye while performing activities, like watching movies, likely to be appealing for children.

Figure 5 shows an extreme configuration of the 3D4Amb image viewer, in which the entire image (sky and balloons) is shown to the lazy eye, while the normal eye receives only the less interesting parts of it (only the sky).

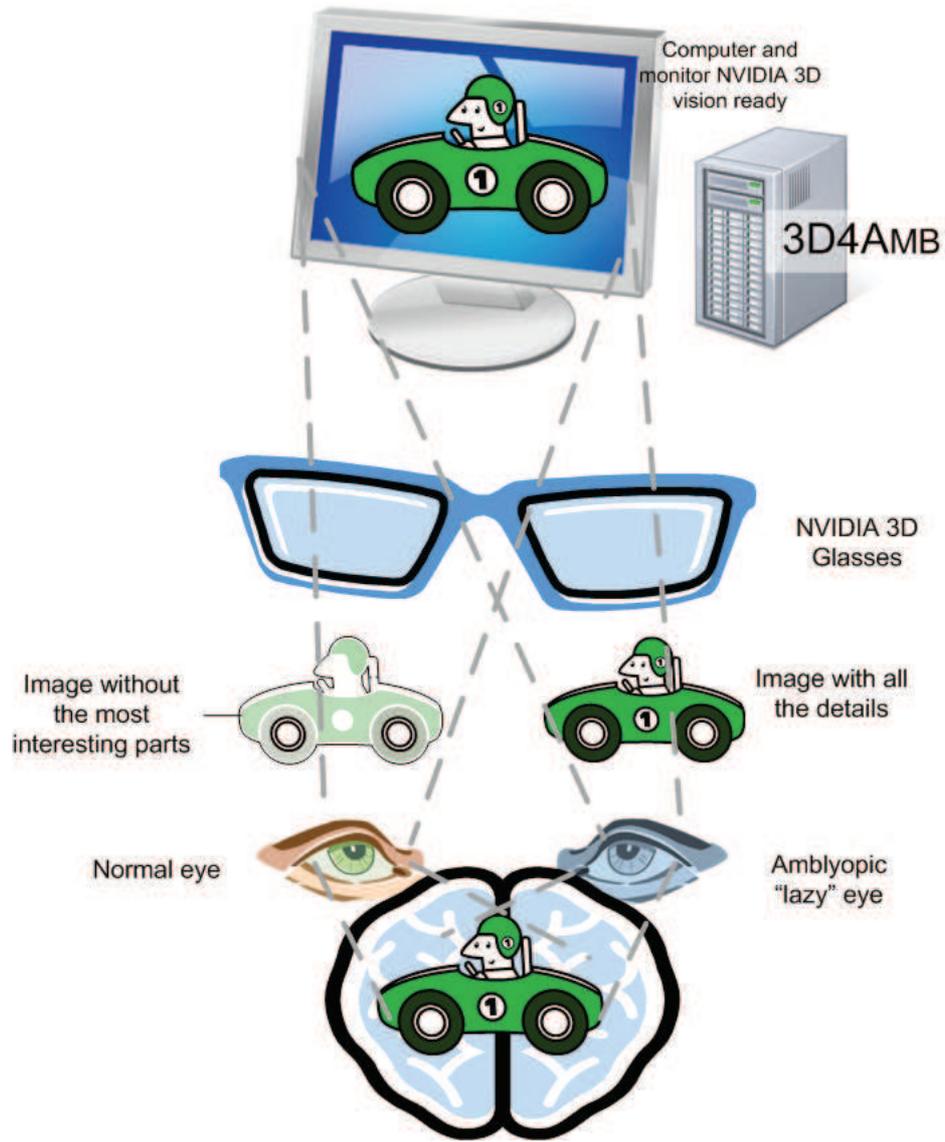

Figure 3: Basic principle of the project

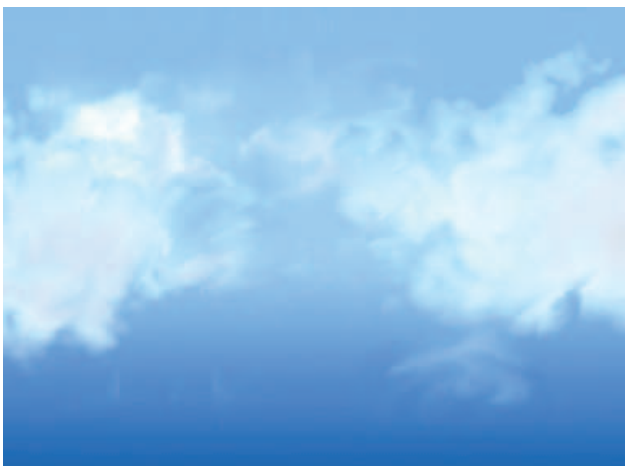

(a) Normal Eye

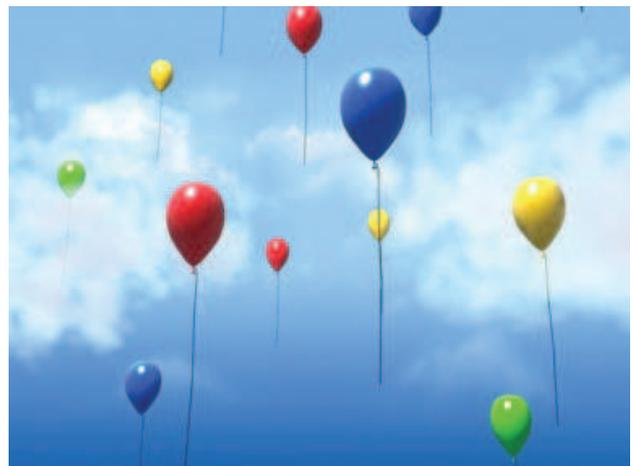

(b) Lazy Eye

Figure 5: Images viewing

*c) Video Games and Exercises.:* While patching is classified as passive method, other treatments which require some activity on the part of the patients are classified as *active*. Active methods are intended to enhance treatment of amblyopia in a number of ways, including increased compliance and attention during the treatment periods (due to activities that are interesting for the patient) and the use of stimuli designed to activate and to encourage connectivity between certain cortical cell types. A good survey and assessment about active treatments and their efficiency can be found in [14].

The most advanced and *active* use of 3D4AMB is the active playing with interactive games or exercises, which will stream binocular images. In this settings, the child plays with a special video game which will exploit the binocular vision to send to the lazy eye all the details while the normal eye will see only a part of the game scene. To successfully complete the game or the exercise the patient must use the information shown to the lazy eye (and possibly fuse it with the information shown to the normal eye). In this way, the amblyotic eye is more stimulated and the fusion encouraged. The game application must continuously monitor the success rate of the game in order to adjust the difficulty based on the real capability of the player. The usage data must be stored to be later inspected by the physician which can check compliance and success, and adjust the type of exercises.

It is well known that video games can be very useful for visual rehabilitation [2]. Classical examples of games found in literature, include PAC-MAN and simple car racing games [16]. As example, we have developed a small space invaders games shown in Figure 6 in which the entire scene is presented to the lazy eye, while the normal eye does not see the spacecraft (red in the Figure) and the shots (yellow in the Figure). The player must hit the invaders (green in the Figure) firing some shots by pressing the space bar on the keyboard (or a fire button of a joystick). Some parts of the background (in yellow in the Figure) are shown to both eyes to facilitate fusion.

## IV. Software architecture

The 3D4AMB software is based on several components which facilitate the development of new applications. Figure 7 shows the architecture stack of 3D4AMB. The system works only (up to now) on a PC with Microsoft Windows® Vista 32/64-bit or Windows 7 32/64-bit Operative systems with the NVIDIA 3D Vision drivers and with Sun Microsystems Java Runtime Environment. Above these, 3D4AMB software is compound of the following modules:

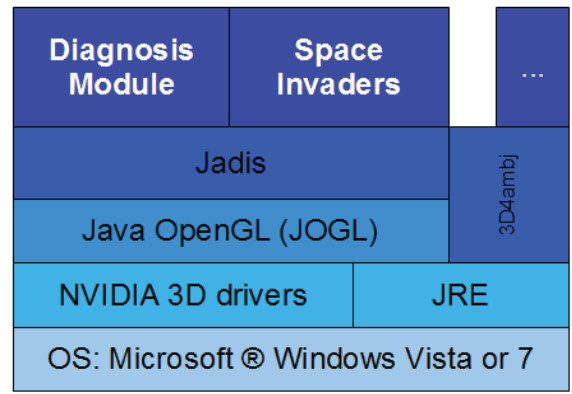

Figure 7: 3D4AMB software stack architecture

- **3D4Ambj Library**: A Java library we have built and now maintain in order to manage data which are shared among all the 3D4AMB applications. For instance, it contains the necessary code to the save a load the XML file containing the data of the patient.
- **Java OpenGL (JOGL)**: is a wrapper library that allows OpenGL to be used in the Java programming language. It is currently an independent open-source project under the BSD license. It is the reference implementation for JSR-231 (Java Bindings for OpenGL).
- **Jadis**: Jadis (Java Advanced Display Infrastructure for Stereo) [1] provides a common interface for displaying Swing GUI components in stereo using either specialized stereo display hardware (e.g. liquid crystal shutter or polarized glasses) or anaglyph display (red/blue glasses) on standard computer displays. An application using this infrastructure will work without modification in either environment, allowing stereo software to reach a wider user base without sacrificing high-quality display on dedicated hardware.
- **Applications**: On the top of our architecture, stay all the applications, like the diagnostic module, the games, and the video and image viewers.

We have chosen Java and JOGL for several reasons. The use of Java permits the reuse of most software we have developed in the past using Java and makes the collaboration with our students easier since they study Java as main programming language. Moreover, Java and JOGL are open standards: 3D4AMB is easily portable over other 3D systems in the future.

## V. Future work - Medical studies

We have started some clinical studies with the help of the Department of Ophthalmology of a local hospital to assess the overall quality and usability of 3D4AMB. We have started already with some healthy patients to find usability problems and identify potential risk factors. We plan in the future to conduct clinical studies to determine the efficiency of 3D4AMB with respect to the classical patching.

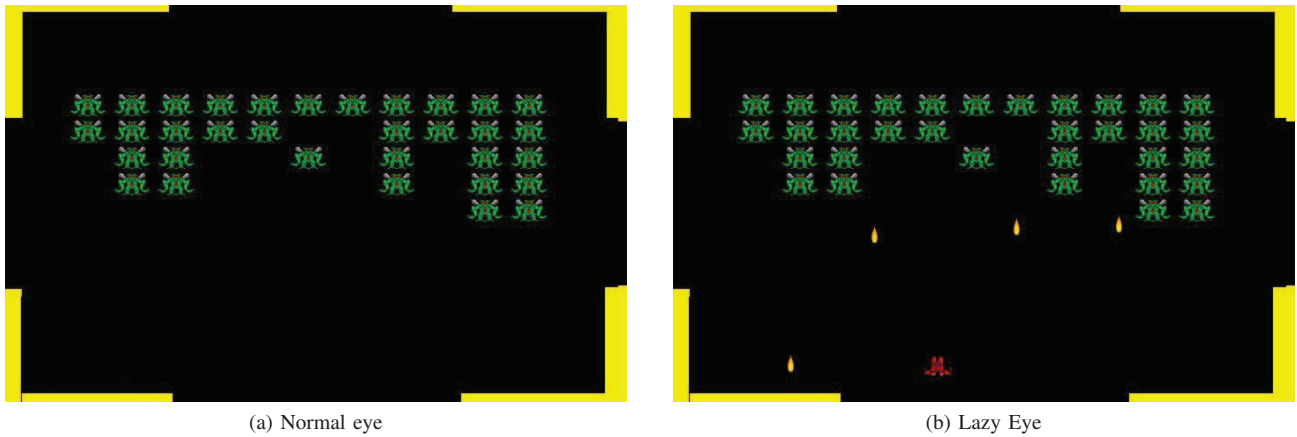

(a) Normal eye

(b) Lazy Eye

Figure 6: Space Invaders Game

## VI. CONCLUSIONS

In this paper we have presented a system, 3D4AMB, for the diagnosis and treatment of amblyopia in young children which is based on 3D vision technologies. The 3D vision is exploited to allow the binocular vision, i.e. to send different images to the normal and the lazy eye, in order to exercise and stimulate the lazy eye and the brain to fuse the images in an unique view. We have presented several use cases supported by prototypes we have developed using Java and JOGL technologies for stereo viewing. The system has been devised with the goal to improve compliance with the treatment. It should not suffer from the non-compliance problems of the classical patching treatment. The children should enjoy the domestic use of 3D4AMB games and exercises, allowing prolonged therapy and the system promise to be inexpensive enough to foster its usage.

## REFERENCES

[1] Java advanced display infrastructure for stereo, http://www.openchannelsoftware.com/projects/jadis. 2008.

[2] R. Achtman, C. Green, and D. Bavelier. Video games as a tool to train visual skills. *Restorative Neurology and Neuroscience*, 26(4-5):435–446, 2008.

[3] M. Cleary. Efficacy of occlusion for strabismic amblyopia: can an optimal duration be identified? *British Journal of Ophthalmology*, 84(6):572–578, 2000.

[4] M. Cleary, A. D. Moody, A. Buchanan, H. Stewart, and G. N. Dutton. Assessment of a computer-based treatment for older amblyopes: the Glasgow Pilot Study. *Eye (Lond)*, 23(1):124–131, Jan 2009.

[5] M. Dixon-Woods, M. Awan, and I. Gottlob. Why is compliance with occlusion therapy for amblyopia so hard? a qualitative study. *Arch Dis Child*, 91(6):491ï¿œ494, June 2006.

[6] R. M. Eastgate, G. D. Griffiths, P. E. Waddingham, A. D. Moody, T. K. H. Butler, S. V. Cobb, I. F. Comaish, S. M. Haworth, R. M. Gregson, I. M. Ash, and S. M. Brown. Modified virtual reality technology for treatment of amblyopia. *Eye*, 20(3):370–374, Apr. 2005.

[7] S. Fateh and C. Speeg. Rebalancing the visual system of people with amblyopia "lazy eye" by using HMD and image enhancement. In R. Shumaker, editor, *HCI (13)*, volume 5622 of *Lecture Notes in Computer Science*, pages 560–565. Springer, 2009.

[8] R. Gregson. Why are we so bad at treating amblyopia? *Eye*, 16(4):461–462, July 2002.

[9] L. Kosikowski and A. Czyzewski. Computer based system for strabismus and amblyopia therapy. In *International Multiconference on Computer Science and Information Technology. 2nd International Symposium on Multimedia - Applications and Processing*, pages 493–496, 2009.

[10] P. S. Moke, A. H. Turpin, R. W. Beck, J. M. Holmes, M. X. Repka, E. E. Birch, R. W. Hertle, R. T. Kraker, J. M. Miller, and C. A. Johnson. Computerized method of visual acuity testing: adaptation of the amblyopia treatment study visual acuity testing protocol. *American Journal of Ophthalmology*, 132(6):903 – 909, 2001.

[11] D. Newsham. Parental non-concordance with occlusion therapy. *British Journal of Ophthalmology*, 84(9):957–962, 2000.

[12] A. Searle, P. Norman, R. Harrad, and K. Vedhara. Psychosocial and clinical determinants of compliance with occlusion therapy for amblyopic children. *Eye*, 16(2):150–155, March 2002.

[13] C. Sik-Lányi and Z. Lányi. Multimedia program for training of vision of children with visual impairment and amblyopia. *JITE*, 2:279–290, 2003.

[14] C. M. Suttle. Active treatments for amblyopia: a review of the methods and evidence base. *Clinical and Experimental Optometry*, 2010.

[15] P. E. Waddingham, T. K. H. Butler, S. V. Cobb, A. D. R. Moody, I. F. Comaish, S. M. Haworth, R. M. Gregson, I. M. Ash, S. M. Brown, R. M. Eastgate, and G. D. Griffiths. Preliminary results from the use of the novel interactive binocular treatment (I-BiT[trade]) system, in the treatment of strabismic and anisometropic amblyopia. *Eye*, 20(3):375–378, Apr. 2006.

[16] P. E. Waddingham, S. V. Cobb, R. M. Eastgate, and R. M. Gregson. Virtual reality for interactive binocular treatment of amblyopia. In *The Sixth International Conference on Disability, Virtual Reality and Associated Technologies*, 2006.